\documentstyle[prl,aps,epsf,multicol]{revtex}
\begin{document}
\draft
\title{Interlayer tunneling in double-layer quantum Hall
  pseudo-ferromagnets} 
\author{L.~Balents$^1$ and L.~Radzihovsky$^2$}
\address{${}^1$~Physics Department, University of California, Santa
Barbara, CA 93106\\
${}^2$~Physics Department, University of Colorado, Boulder, CO 80309}
\date{Date:\today}
\maketitle

\begin{abstract}
  We show that the interlayer tunneling I--V in double-layer quantum
  Hall states displays a rich behavior which depends on the relative
  magnitude of sample size, voltage length scale, current screening,
  disorder and thermal lengths. For weak tunneling, we predict a
  negative differential conductance of a power-law shape crossing over
  to a sharp zero-bias peak.  An in-plane magnetic field splits this
  zero-bias peak, leading instead to a ``derivative'' feature at
  $V_B(B_{||})=2\pi\hbar v B_{||}d/e\phi_0$, which gives a direct measure of
  the dispersion of the Goldstone mode corresponding to the
  spontaneous symmetry breaking of the double-layer Hall state.
\end{abstract}

\begin{multicols}{2}
\narrowtext 

Through a series of experimental and theoretical papers, it is now
well-established that a bilayer of two-dimensional (2d) electron gases
(2DEGs), when the layers are sufficiently close, exhibits an
incompressible Quantum Hall (QH) state at a total filling fraction
$\nu=1$,~\cite{eisenstein92,wen_zee,murphy}, even when interlayer
tunneling is negligible.~\cite{yang_moon}.  The QH state is stabilized
by the Coulomb exchange interaction, which leads to {\em spontaneous}
interlayer phase coherence, with the layer phase difference $\phi$,
the ``phason'', as the corresponding Goldstone mode. Because the layer
index may also be viewed as a pseudo-spin-$1/2$ degree of freedom,
this state is also known as the QH FerroMagnet (QHFM), and the phasons
as (pseudo-)magnons.

The first {\sl direct} experimental evidence for this mode was
obtained recently by Spielman, et al.\cite{eisenstein00}.  For
bilayers with small separations $d$, they observed a giant and
spectacularly narrow zero bias peak in the inter-layer tunneling
conductance at low $T$, a feature that naturally emerges from our
study of the QHFM. We find rich variations of the {\sl non-linear}
tunneling conductance $G(V) = dI/dV$ as a function of interlayer
voltage $V$, tunneling strength $\Delta_0$, disorder, temperature, and
applied parallel magnetic field $B_\parallel$.  Two particularly
appealing and testable predictions of our theory are: (1) a
finite-bias feature in $G(V)$ at a voltage $V_B = \hbar \omega(Q= 2\pi
B_\parallel d/\phi_0)$ which tracks the collective mode dispersion
relation $\omega(Q)$ as a function of $B_\parallel$ ($\phi_0 = hc/e$)
and (2) the appearance of a power-law {\sl negative differential
resistance} at higher bias in cleaner samples.  For clarity, below we
first summarize different regimes and results, and then give a brief
sketch of the calculations.




We focus here on voltages $eV$ below the QH charge gap $\Omega$, where
the only available bulk excitations are phasons, described in the
absence of impurities by the Lagrangian
\begin{equation}
\hspace{-0.15in} {\cal L} = 
{\rho_s\over 2}\left[ {1 \over v^2}(\partial_t \phi)^2 -
    |\bbox{\nabla}\phi|^2\right] - {\Delta_0\over2\pi\ell^2} 
     \cos\left[\phi\! -\! \omega t\! -\! Q x\right],
\label{Lagrangian}
\end{equation}
where $v$ is the pseudo-magnon ``sound'' velocity, $\rho_s$ is the
phase stiffness,\cite{yang_moon}\ $\ell = \sqrt{\hbar c/eB}$ is the
magnetic length, $Q$ is the in-plane ``magnetic wavevector'' $Q=2 \pi
B_{||}d/\phi_0$ due to applied ${\bf B}_{||}=B_{||}\hat{\bf y}$, and
$\omega = eV/\hbar$.  The only other low-energy modes are {\sl edge
  states}, which carry the in-phase, in-plane current at low
temperatures.

For weak tunneling, characterized by a ``naive'' dimensionless
coupling $\delta_0\equiv\Delta/\rho_s\ll 1$, the tunneling current
density $J(V)$ can be computed in a controlled perturbation theory
(but see below).  There are three interesting sub-cases.  First, for
the pure system, we find
\begin{equation}
J(V)= \cases{J_0\delta_0\left({{\cal E}_\ell\over k_B T}\right)^2
     \left({k_B T\over e V}\right)^{2-\eta} \mbox{sgn}(V), &
     $e V \ll k_BT\;$, \cr 
     \sim \exp(-e V/{k_B T}) & $e V \gg k_B T$.}
\label{J0perturbative}
\end{equation}
where, $\Delta\equiv C_Q \Delta_0$ is the tunneling amplitude downward
renormalized by quantum fluctuations, with $C_Q < 1$ a constant
defined below. In above $J_0\equiv e\Delta/\hbar2\pi\ell^2$ is the
tunneling current density scale, and ${\cal E}_\ell\equiv \hbar
v/\ell$ is the ultra-violet energy cutoff associated with the magnetic
length $\ell$ (presumably $O(\Omega)$ for $d/\ell$ of $O(1)$),
$\eta=k_B T/2\pi\rho_s$ is the familiar Kosterlitz-Thouless (KT)
exponent\cite{KT} controlling the asymptotic quasi-long range order
(qlro) of the double-layer QH state.  The exponential suppression at
high-bias is a kinematic consequence of the lack of zero momentum
excitations in this energy range.

Note that the above I--V displays {\sl negative differential
  resistance}, and a peak in $J(V)$ as $V\rightarrow 0^+$.  The
physics of this peak is the appearance of {\sl excitons}, or
electron-hole pairs from opposite layers as sharply defined low energy
excitations in the QHFM, whose spectral function is directly probed by
the tunneling conductance.


Second, sufficiently strong impurities modify the transport (and other
aspects of the physics), introducing a number of new length scales.
We discuss their effects primarily in the regime in which tunneling
remains perturbative (but see below).  If disorder is not too weak
(weak impurities do not significantly modify the behavior),
significant quenched-in {\sl static} deformations in $\phi$ appear
beyond a length $\xi$, removing the constraint of in-plane momentum
conservation.  Then for $eV \ll k_B T$,
\begin{equation}
  J_d(V) \approx J_0\delta_0 \left({\xi\over\ell}\right)^2 
    \left({{eV} \over {k_B T}}\right)^\alpha {\rm
    sgn}(V), \label{dirty}
\end{equation}
where naively $\alpha=\eta$, but preliminary investigations suggest
that at low temperature quantum effects can lead to an enhanced
$\alpha<1$.\cite{BRunpublished}\  At large bias, $eV \gg \hbar v/\xi
\gg k_B T$, the current decays exponentially ($I_d \sim \exp(-
eV\xi/\hbar v)$). 

Third, in the presence of an additional weak $B_{||}$ field the
zero-bias peak of Eq.\ref{J0perturbative} is split by $V_B=\hbar v
Q/e$,
\begin{equation}
J_Q(V)= J_0{\delta_0{\cal E}_\ell\over\ell Q k_{\rm B}T}\sum_{s=\pm1}
 s\left|{k_B T\over e V- s Q v\hbar}\right|^{1-\eta}.
\label{JQperturbative}
\end{equation}
Eq.~\ref{JQperturbative}\ is valid for weak impurities, $Q\xi \gg 1$,
except in a narrow region $|eV - \hbar v Q| \lesssim \hbar v/\xi$, in
which the singular peak is rounded by disorder and strong-coupling physics.  
Consequently we find
that the location of the tunneling conductance feature (which is a
peak for $Q=0$ and is of a ``derivative'' shape form for $Q>0$) maps
out the wavevector ($Q$) dispersion of the Goldstone mode associated
with the transition to the double-layer QH state and with the broken
U(1) symmetry.   

The naive application of the above perturbative results, $I_{\rm uns}
= J(V) L^2$ (for a square $L\times L$ sample), can break down for
several reasons.  First, in the experimentally relevant geometry with
contacts at opposite sample ends, there is necessarily a variation of
the chemical potential in the plane of the 2DEG in order to maintain
the in-plane current.  Thus the $V$ in
Eqs.~\ref{J0perturbative}-\ref{dirty}\ becomes spatially dependent,
and moreover, for sufficiently large samples most of the voltage drop
is in-plane.  Second, and more interestingly, the perturbation theory
itself breaks down at low bias.  This occurs once the voltage scale
$L_V = \hbar v/eV$ becomes longer than a ``penetration depth''
$\lambda(V,T)$ (provided $\lambda<L$) defined below.  Physically,
$\lambda$ represents a screening length beyond which the in-plane
out-of-phase currents become too large to support a uniform tunneling
current in the absence of another channel (e.g. contacts or edge
states) to carry away the transferred charge.

In the pure case, we find the {\sl non-equilibrium} penetration depth
$\lambda(V,T)\approx\lambda_0(\lambda_0/L_V) (k_B T/e V)^{\eta/2}$.
A rather idealized experimental geometry for which this prediction may
directly apply is for two parallel contacts to the {\sl same side}
(but different layer) of the bilayer.  Then the transport occurs only
within a distance $\lambda$ of the contacts, within which the
tunneling current {\em density} is comparable to that in
Eq.~\ref{J0perturbative} and therefore the current is given by
$I_{scr}=L\lambda(V,T) J(V)$.  Note that in the ``perturbative''
high-bias regime $e V/k_B T\gg\left[\delta_0({\cal
E}_\ell/k_BT)\right]^{2/(4-\eta)}$ (equivalently defined by
$L_V<\lambda(V,T)$), $\lambda(V,T)$ is always longer than
$\lambda_{\rm equil}(T)=\lambda_0(\lambda_0 k_B T/\hbar
v)^{\eta/(4-\eta)}$ (a length that one would have naively expected
from an {\em equilibrium} calculation), but reduces to it for a lower
bias. For large tunneling, $\delta_0>1$, the penetration depth is
instead given by a significantly shorter strong coupling expression
$\lambda_0=\ell\sqrt{2\pi\rho_s/\Delta}$.  For the more experimentally
relevant geometry with contacts on opposite sides of the sample, the
tunneling current must be carried primarily along the sample {\sl
edges}, as in the quantum Hall regime only the edge states can carry
the {\sl symmetric} current needed by charge conservation.
With $\xi<\infty$, screening is reduced and the corresponding
``dirty'' penetration depth becomes
$\lambda_d(V)\sim\lambda_0(\lambda_0/\xi)(k_BT/eV)^{\eta/2}=(L_V/
\xi)\lambda(V,T)$, provided the self-consistent condition
$\xi<\lambda$ and new physics does not intervene at lengths smaller
than $\lambda_d$.

Additional randomness-induced phenomena can also occur at scales
beyond $\xi$, which corresponds to the length at which unbound {\sl quenched}
vortices are present.  Although
at low temperatures vortex motion is exponentially suppressed due to
the weakness of thermal activation and quantum tunneling, because it
is nevertheless finite, it leads to {\em temporal} fluctuations of the
interlayer phase $\phi$ (which we may crudely call ``decoherence'')
for times longer than the exponentially long time scale $\tau_{\rm
vort}$ associated with vortex motion.  Therefore, provided any unbound
vortices are induced, we expect an exponentially diverging zero-bias
conductance peak $dI/dV|_{V=0} \sim \tau_{\rm
vort}^{1-\eta}|\Delta|^2$ of width 
$\hbar/\tau_{\rm vort}$.  Ultimately, this divergence with reducing
temperature is cut off by mesoscopic effects in finite samples --
phase ``diffusion'' due to energy and current exchange with the leads,
tunneling contributions from edge modes, etc. -- and by effects
non-perturbative in the tunneling, to which we now turn.   Finally,
for strong enough disorder even the gauge glass order may be destroyed
at $T=0$.  The nature of the resulting state and the transition to it
remain unresolved.

All the above results assume that $J(V)$ (at least in a large region
near the contact) is perturbative in the tunneling $\Delta$.  At
sufficiently low $V$ and $T$, however, this perturbation theory breaks
down.  We first consider the clean limit, in which the ``naive''
dimensionless coupling that controls this expansion is
$\delta(T)\equiv\delta_0({\cal E}_\ell/k_BT)^2$.  The true expansion
parameter is $\delta_{\mbox{eff}}=\delta(k_B T/eV)^{(4-\eta)/2}$, as
can be seen from standard sine-Gordon RG analysis.  In the strong
interlayer tunneling regime, defined by $e V/k_B
T<\delta^{2/(4-\eta)}$, currents are screened in the bulk on the scale
$\lambda$.  For contacts on the same sample edge, tunnel currents run
only near this edge and the I--V curve measures the conductance of the
leads.  For contacts on opposite sample edges, the in-plane current
can be carried only by the edge states, and the model must be extended
to incorporate these degrees of freedom.  However, on general grounds
(see below) we expect that the power-law I--V, displaying negative
differential resistance, below this strong coupling voltage,
generically crosses over to a linear I--V, corresponding to a large,
narrow conductance peak, as observed in
experiments.\cite{eisenstein00}\

We now present the highlights of calculations that lead to the
tunneling results summarized above, defering most of the
details\cite{BRunpublished} to a future publication.  We focus on the
tunneling current density $J =
{e\Delta_0\over2\pi\ell^2\hbar}\langle\sin\left[\phi - \omega t - Q
x\right]\rangle$, where the angular brackets denote an expectation
value in the non-equilibrium steady state at time $t$ averaged over a
thermal ensemble of initial conditions in the far past.  Note that, in
the mesoscopic regime in which $\hbar v/L > eV, k_B T$, it is crucial
to include the degrees of freedom of the contacts in the
non-equilibrium ensemble.

Standard time-dependent perturbation theory, to leading nontrivial
order in $\Delta_0$ gives
the {\sl current density}
\begin{equation}
  J = J_0 {\Delta_0\over2\pi\ell^2\hbar}
  {\rm Re}\, \int\! d^2{\bf r} \int_0^\infty \!\! dt\, e^{i(Qx - \omega
    t) - 0^+ t} C({\bf r},t),
\label{J2}
\end{equation}
where $C({\bf r},t)=\theta(t)\left\langle \left[e^{i\phi({\bf
r},t)},e^{-i\phi({\bf 0},0)}\right]\right\rangle_0$, calculated using
the unperturbed Hamiltonian with $\Delta_0=0$.

A straightforward calculation gives
\begin{equation}
  C({\bf r},t) = C_Q\ e^{-G_{\rm th}({\bf r},t)} \left(e^{-G_{\rm Q}({\bf
          r},t)} -  {\rm c.c.} \right),
\label{C2}
\end{equation}
where $C_Q=\left[\cosh(\pi{\cal E}_\ell/2k_B T)\right]^{-2\eta}\approx
e^{-\hbar v/\ell\rho_s}$, and
\begin{mathletters}
\begin{eqnarray}
  G_{\rm th}({\bf r},t) & = & {v\hbar \over \rho_s} \int\! {{d^2 q} 
    \over {8\pi^2 q}} \left| e^{i{\bf q \cdot r} - i v q t}\! - \!
    1\right|^2 n_{\rm \scriptscriptstyle B}(\hbar v q), \label{Gth}\\
  G_Q({\bf r},t) & = &{v\hbar \over {4\pi\rho_s}} \left[{{\theta(r-vt)} \over
      \sqrt{r^2-v^2 t^2}} - i {{\theta(vt-r)} \over
      \sqrt{v^2 t^2-r^2}}\right],\label{Gq}
\end{eqnarray}
\end{mathletters}
are the ``thermal'' and ``quantum'' propagators and $n_B(\epsilon) =
[e^{\epsilon/k_BT}-1]^{-1}$.  Eqs.~\ref{Gth}-\ref{Gq}\
neglect the quantization of the total charge across the bilayer
capacitor, and therefore misses Coulomb blockade effects, which are
small since for large systems, the charging energy $e^2/C = \hbar^2
v^2/\rho_s L^2$ is much less than the ballistic ``level spacing''
$\hbar v/L$.  Aside from the presence of $G_{th}$,
Eqs.~\ref{J2}-\ref{Gq}\ are identical to the result of {\sl classical}
perturbation theory in the equation of motion following from
Eq.~\ref{Lagrangian}.

$G_Q({\bf r},t)$ is independent of $T$ and hence reflects the quantum
dynamics of the system (for $\Delta=0$).  By inspection of
Eq.~\ref{C2}, only when $G_Q({\bf r},t)$ contains a non-zero {\sl
imaginary} part is $C({\bf r},t)$ non-zero.  Therefore the tunneling
current is non-zero only for $vt>r$, reflecting causality.
Since we are interested in low energy scales, ($eV,\hbar v Q,k_B T \ll
{\cal E}_\ell$) 
we expect the tunneling current to be dominated by the large $r,t$
behavior of $C(r,t)$, with $\rho_s t \gtrsim \rho_s r/v \gg \hbar$.
Then, except in a very narrow range (which contributes negligibly to
the current) in which $0 < vt-r \ll \hbar v/\rho_s$, $G_Q \ll 1$, leading to
\begin{equation}
  C(r,t)\approx - {iv\hbar \over {2\pi \rho_s}} C_Q e^{-G_{th}(r,t)}
  {\theta(vt\!-\!r) \over \sqrt{v^2t^2-r^2}} .
  \label{C4}
\end{equation}

The effect of thermal fluctuations are encoded in $G_{\rm th}(r,t)$.
The tunneling current contains two contributions.  Modes at time
scales smaller than the thermal time, $\tau_T=\hbar/k_{\rm B}T$ give
$T=0$ contributions with $e^{-G_{\rm th}}\approx 1$.  In contrast, for
modes with $vt > r \gg v\tau_T$, $G_{\rm th}(r,t)\gg 1$, with dominant
contribution coming from thermally activated modes with $\hbar v q \ll
k_{\rm B}T$, for which $n_{\rm \scriptscriptstyle B}(\epsilon)\approx
k_{\rm B}T/\epsilon$ and the upper cutoff on $q$ given by $k_{\rm
B}T/v\hbar$.  Computing the integral over ${\bf q}$ in Eq.\ref{Gth}
under such approximation we find
\begin{equation}
  e^{-G_{\rm th}(r,t)} \approx   \tanh\left\{ \left[{{v\tau_T} \over
      {vt+\sqrt{v^2t^2-r^2}}}\right]^\eta  \right\}.
\label{Gth2} 
\end{equation}
Eq.~\ref{Gth2}\ must be taken together with an implicit causality
constraint $r<vt$, (see Eq.\ref{C4}), so that the argument of $\tanh$
is always real and of order $(\tau_T/t)^\eta$ for $t \gg \tau_T$.  The
well-known KT theory\cite{KT} restricts the applicability of
Eq.~\ref{Gth2}\ to $\eta<1/4$, for which the qlro'ed phase is stable
to thermally-induced vortices.  Finally, for bilayers which are not
too large, an IR cut-off should be introduced to restrict $r<L$.  The
behavior of $C(r,t)$ for $t>L/v$, is strongly modified by the presence
of the contacts even at $T=0$.  This provides an additional source of
``decoherence'' to smooth the singularities of the low-bias tunneling
conductance for $e V< \hbar v/L$.

We now apply these results for the Green's functions to the analysis
of the tunneling current in a macroscopic, clean system, with $L>L_V$,
where the results are independent of the contact physics. On physical
ground we expect that for the most parts our predictions will also
extend into the mesoscopic regime, $L<L_V$, with the mesoscopic modes
(not fully accounted for by the $\phi$-only model) only contributing a
featureless background to the tunneling I--V.  Combining Eqs.~\ref{C4}
and \ref{Gth2} with Eq.\ref{J2}, we arrive at a naive perturbative
expression for the tunneling current {\sl density} in an infinite
sample,
\begin{eqnarray}
J_Q(V) & \approx & {J_0\Delta\over4\pi\rho_s}{C_Qv\over2\pi\ell^2}
  \int_0^\infty
  \!dt\sin(e V t/\hbar) \int_0^{vt} \!\!\!\!\! dr r {{J_0(Qr)} \over
  \sqrt{v^2 t^2 - r^2}}\nonumber \\
  & & \times \tanh\left\{\left[ {v\tau_T\over {(vt+\sqrt{v^2 t^2-r^2})}}
  \right]^\eta\right\}, \label{IhighT}
\end{eqnarray}
which readily leads to Eqs.~\ref{J0perturbative},\ref{JQperturbative}.

Based on an equilibrium RG analysis, these expressions would naively
appear to be valid in the thermodynamic limit $L\rightarrow \infty$,
provided $e V/k_B T\gg\delta^{2/(4-\eta)}$.  The true regime of
validity is, however, more restrictive.  In particular,
Eq.~\ref{IhighT}\ predicts a non-zero steady-state current {\sl
density} in the infinite system.  This would require that either the
in-plane (out-of-phase) current or the charge density across the
bilayer ``capacitor'' be {\sl infinite}.  The apparent violation of
charge conservation signals a breakdown of the perturbative results
for $L>\lambda$.  Technically it occurs because even large bias $V$ is
not sufficient to cut off all long scale perturbative
divergences. However, the remaining divergences are weaker and
therefore the length scale $\lambda(V,T)$ over which perturbation
theory is valid is {\sl longer} than the equilibrium screening length
$\lambda_{\rm equil}(T)=\lambda_0(\lambda_0 k_B T/\hbar
v)^{\eta/(4-\eta)}$ relevant for low bias.  Physically, $\lambda(V)$
may be estimated by considering a steady state, for which the
continuity equation reduces to ${e\over\hbar}\rho_s \nabla^2 \phi =
J(V)$, since the in-plane pseudo-spin supercurrent density $j_s =
{e\over\hbar}\rho_s{\bbox\nabla}\phi$.  The naive, {\em uniform}
$J(V)$ can be maintained only over lengths such that the mean $\phi$
grows by an amount $\ll 1$.  Thus
$\lambda(V)\approx\sqrt{e\rho_s/\hbar J(V)}$, which leads to the
result quoted in the introduction.  To support this interpretation, we
note that $\lambda(V)$ reduces to $\lambda_{\rm equil}(T)$ if the
naive boundary of perturbation theory, $e V/k_B
T\approx\delta^{2/(4-\eta)}$, is approached. For the case of both
contacts to the same sample edge ($x=0$) and $L>\lambda(V)$, we have
verified perturbatively in $\Delta$ that the {\sl non-uniform} current
density profile is $J(x,V) = J(V)F(x/\lambda(V))$, with
$F(\chi)\rightarrow 0$ for $\chi \rightarrow \infty$ and $F(\chi) \sim
1 - O(\chi^2)$.

The most significant effect of weak disorder is an effective random
vector potential ${\bf a}({\bf r})$ (${\bbox\nabla}\phi \rightarrow
{\bbox\nabla}\phi - {\bf a}({\bf r})$) in
Eq.~\ref{Lagrangian}.\cite{stern99,BRunpublished} This arises from the
binding of charge to flux in the QH regime, but has {\sl short-range}
correlations since it is a physical field.  Thus in the zero tunneling
limit, for weak disorder we arrive at a $2+1$-dimensional XY model
with random vector potential (XYRVP).  The XYRVP is known to have a
stable low $T$ phase with qlro and no unbound vortices below a
critical strength of disorder.\cite{nattermann}\ At some threshold
strength of disorder, vortices will be induced, and the system is
effectively in a {\sl gauge glass} regime.  The vortices introduce an
asymptotically exponential decay in $r$ and $t$
\begin{equation}
C({\bf r},t) \rightarrow C({\bf r},t) e^{-r/\xi} e^{-t/\tau_{\rm vort}}.
\label{decay}
\end{equation}
The characteristic length $\xi$ diverges as the threshold disorder
strength is approached from above, and is finite above it, even at
$T=0$.  Vortex localization implies that $\tau_{\rm vort}$ diverges as
$T\rightarrow 0$, so $\tau_{\rm vort} \gg \xi/v$.  Indeed, there is
very likely an intermediate regime $\xi/v \ll t \ll \tau_{\rm vort}$
over which $C({\bf r},t)\sim t^{-(1+\alpha)}$, with no rigorous theory
for $\alpha$, but strong arguments for $\alpha<1$.\cite{BRunpublished}

Below the threshold
disorder strength, such that $\xi=\infty$, the XYRVP model may be used
to calculate the $C({\bf r},t)$.  This gives a weak modification of
the $\eta$ exponent by quenched impurities but leaves the results
otherwise unchanged.  Above threshold, the effects of disorder are
stronger.  Eq.~\ref{decay}\ may then be used,
leading to the perturbative result in Eq.~\ref{dirty}.  As for the
pure case, such non-zero $J_d(V)$ cannot persist in the thermodynamic
limit.  Similar considerations give the result for the dirty screening
length quoted in the introduction.  

Two different behaviors are expected for weak disorder depending upon
sample size.  If $L<\lambda$, the low $V$ divergence of $J(V)$ is cut
off by decoherence as phasons scatter off the leads.  For large
samples ($L>\lambda$), the perturbative results are only valid when
$L_V<\lambda(V)$, with system in a strong coupling, nonperturbative in
$\Delta$ regime for lower bias.  In this limit, we expect a {\sl
linear} response on general hydrodynamic grounds.  In the presence of
tunneling and impurities, the only operative conservation laws are for
total charge and energy.  Hence, the low-frequency, long-wavelength
equations of motion (in addition to that for $\phi$) must take the form
\begin{equation}
\partial_t n_i = - \gamma_{ij} \mu_j - \partial_\mu j^\mu_i, \label{hydro}
\end{equation}
where $n_i$ and $\mu_i$ are, respectively, the density and chemical
potential in layer $i$, $\gamma_{ij} = \gamma (2\delta_{ij} - 1)$,
$j^\mu_i = \sigma^{\mu\nu}_{ij} \partial_\nu \mu_j$ are the in-plane
currents, and repeated indices are summed.  Note that Eq.~\ref{hydro}\
combined with the QH condition $\sigma_{ii}^{xx} \ll
\sigma_{ii}^{xy}$, gives a classical description of edge transport.
The coefficient $\gamma$ thereby provides a hydrodynamic definition of
the tunneling ``conductivity'', closely related to the true measured
inter-layer conductance.  For {\sl zero} tunneling, $n_1 -n_2$ is
conserved, and model F (xy-ferromagnet) hydrodynamics should
apply.\cite{Halperin_Hohenberg} Hence, for small $\Delta$, we expect
that $\gamma(\Delta)$ is determined by the {\sl classical} crossover between
these two limits, which to our knowledge is not understood.  A scaling
approach which assumes a direct crossover suggests (for weak disorder)
$\gamma\sim (k_B T/\hbar)({\cal E}_\ell/k_B
T)^{4/(4-\eta)}(\Delta/\rho_s)^{2/(4-\eta)}$.  At low $T$ (especially
for stronger disorder), where the renormalized $\Delta$ becomes large
already in the quantum regime, the crossover to the hydrodynamic limit
takes place in at least two stages, and is more difficult to estimate.
The general picture which emerges, however, is that tunneling currents
are screened to sample boundaries.  Even at low $T$, this is expected
to lead to {\em linear} transport as $V\rightarrow 0$, and for
particular parameters this limit can be analytically studied using
Eq.~\ref{hydro}.

Although a detailed analysis of the strong coupling crossover is
difficult, for very weak disorder, the above scaling considerations
are corroborated by a complementary {\sl semiclassical} analysis,
which treats the kinetic and tunneling terms on equal footing.
As a test of principal, we have considered in detail an idealized pure
system with parallel side contacts. We find that the current is then
carried by a ``half soliton'' pinned to the contacts.  Dissipation
arises due to a non-local Caldeira-Leggett coupling arising from the
leads.  This gives a linear I--V curve at low bias with a conductance
that has the same scaling form and magnitude as predicted by the
hydrodynamic approach above.

L.B. was supported by the NSF--DMR--9985255, L.R. acknowledges support
by the NSF DMR-9625111, and by the Sloan and Packard Foundations.

\end{multicols}
\end{document}